\documentclass[superscriptaddress,aps,prl,twocolumn,showpacs,preprintnumbers,amsmath,amssymb,floatfix,groupedaddress]{revtex4}
\usepackage{graphics}
\usepackage{graphicx}
\usepackage{amssymb}
\usepackage{bm}
\newcommand{\bra}[1]{\langle#1|}
\newcommand{\ket}[1]{|#1\rangle}

\newcommand{\out}[2]{\ket{#1}\bra{#2}}
\newcommand{\expect}[1]{\langle#1\rangle}
\newcommand{\rmd}{\mathrm{d}}
\newcommand{\mi}{\mathrm{i}}

\begin{document}
\title{Mapping broadband single-photon wavepackets into an atomic memory}
\author{J Nunn}
\email{j.nunn1@physics.ox.ac.uk} \affiliation{Clarendon Laboratory, Oxford University, Parks Road, Oxford OX1 3PU, UK}
\author{I A Walmsley}
\affiliation{Clarendon Laboratory, Oxford University, Parks Road, Oxford OX1 3PU, UK}
\author{M G Raymer}
\affiliation{Oregon Center for Optics and Department of Physics University of Oregon, Eugene OR 97403, USA}
\author{K Surmacz}
\affiliation{Clarendon Laboratory, Oxford University, Parks Road, Oxford OX1 3PU, UK}
\author{F C Waldermann}
\affiliation{Clarendon Laboratory, Oxford University, Parks Road, Oxford OX1 3PU, UK}
\author{Z Wang}
\affiliation{Clarendon Laboratory, Oxford University, Parks Road, Oxford OX1 3PU, UK}
\author{D Jaksch}
\affiliation{Clarendon Laboratory, Oxford University, Parks Road, Oxford OX1 3PU, UK}
\begin{abstract}
We analyze a quantum optical memory based on the off-resonant Raman interaction of a single broadband photon, copropagating with a classical control pulse, with an atomic ensemble. The conditions under which the memory can perform optimally are found, by means of a `universal' mode decomposition. This enables the memory efficiency to be specified in terms of a single parameter, and the control field pulse shape to be determined via a simple nonlinear scaling. We apply the same decomposition to determine the optimal configurations for read-out.
\end{abstract}
\pacs{03.67.-a, 03.67.Dd, 03.67.Hk, 03.67.Lx} \maketitle
\paragraph{Introduction} The conversion of optical information between photons and atoms forms the basis of a quantum interface that is a critical component of quantum communications networks and distributed quantum computers \cite{cryptography,longdistance,repeater}. Such quantum information processing schemes require the ability both to move quantum information between nodes of a network, and to store the information. An important class of quantum optical memories is based on the interaction of individual photons with atomic ensembles, in which the information is transferred coherently from a single photon to a collective excitation of the atoms. If active feedback \cite{polzikreview} is not used, these memories are generally optically dense, with the density controlled dynamically using an ancillary field \cite{EIT1,EIT2,crib}. 

\begin{figure}[h]
\begin{center}
\includegraphics[height=3.5cm]{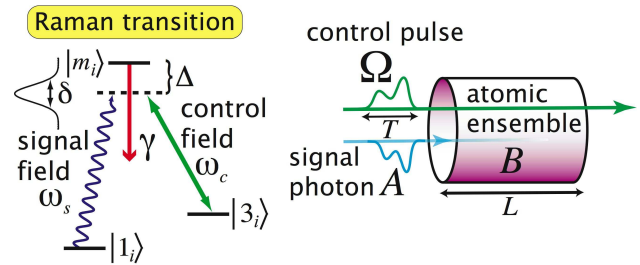}
\caption{Left: the level structure of the $i^{\mathrm{th}}$ atom comprising a quantum memory for broadband photons, with
bandwidth $\delta$. Right: a schematic of the read-in process for the quantum memory.} \label{Figure1}
\end{center}
\end{figure}

In this paper we analyze a prototypical memory based on the off-resonant Raman interaction of a classical pulsed \emph{control} field and a broadband \emph{signal} photon in an atomic medium (see Fig.~\ref{Figure1}). The temporal structure of the signal photon is transferred by the control to a long-lived collective atomic excitation \cite{EIT1}, or \emph{spin wave}. This differs qualitatively from previous narrowband schemes \cite{polzikscheme} which demonstrated that quadrature squeezing could be transferred from optical fields to a collective atomic spin. We show how proper shaping of the control field allows the mapping of an input wavepacket of arbitrary temporal shape to an output wavepacket of a potentially different temporal shape. The dynamics are closely related to those investigated in proposals for entanglement generation via spontaneous \cite{3Dpaper,longdistance} and, more recently, stimulated \cite{raymer2,Wojtek} Stokes scattering. However the photon storage process is distinct from these, in that it exhibits an explicit time reversal symmetry; evinced by its fundamental mode structure. Spontaneous emission is suppressed as long as the excited state remains empty. Lossless unitary storage of broadband single photons -- such as are commonly used in cryptographic and teleportation experiments \cite{cryptography,teleportation} -- can therefore be implemented by detuning sufficiently from resonance \footnote{Shortly after our work was completed, Gorshkov \emph{et al.} presented a paper, quant-ph/0604037 (2006), claiming that the storage of broadband photons on resonance is possible given sufficient atomic density and unlimited control pulse energy.}. Departure from resonance makes our scheme robust against inhomogeneities in the ensemble, so that solid state absorbers (e.g. semiconductor charge quantum dots \cite{sham}) could be substituted for the atoms. In addition, changing the detuning of the control pulse between storage and retrieval allows for control over the frequency of the output state. In the following we consider propagation in the one dimensional limit; a fully three dimensional model will be considered elsewhere.
\paragraph{Model}
The signal and control fields are Raman resonant, with center frequencies $\omega_s$, $\omega_c$, respectively. The classical control at time $t$ and position $z$ is represented by the Rabi frequency $\Omega(\tau)$, where $\tau\equiv t-z/c$ is the \emph{local time}. The signal and spin wave amplitudes are
described by the slowly varying annihilation operators $A(\tau,z)$, and $B(\tau,z)$, respectively \footnote{We work in the linear regime where the operator nature of the variables is not manifest; we retain the operator notation to emphasise that our results hold for non-classical input states.}. The spin wave is a
collective coherence of the form $B(\tau,z)\propto \sum_\beta \out{1_\beta}{3_\beta}e^{-\mi(\omega_s-\omega_c)\tau}$, where the index $\beta$ runs over all
atoms with position $z$ \cite{raymer2}. If the common detuning $\Delta$ of the signal and control
pulses from single photon resonance is much larger than the signal bandwidth $\delta$, the control Rabi frequency $\Omega$, and the control bandwidth, the excited state
$\ket{m_i}$ can be adiabatically eliminated. If the ensemble is prepared in the collective state
$\ket{0}\equiv \bigotimes_i \ket{1_i}$, and if the population of the metastable state $\ket{3_i}$ is assumed to
remain negligible, a linear theory can be used. The Maxwell-Bloch equations, in the slowly varying envelope
approximation, are then found to be \cite{raymer2}\cite{Wojtek}
\begin{eqnarray}
\label{MBs2.1}
\left[\partial_\tau -\mi |\Omega(\tau)|^2/\Gamma\right]B(\tau,z)=&-\kappa^* \Omega^*(\tau)A(\tau,z)/\Gamma;\\
\label{MBs2.2}
\left[\partial_z -\mi |\kappa|^2/\Gamma\right]A(\tau,z)=&\kappa \Omega(\tau)B(\tau,z)/\Gamma,
\end{eqnarray}
where $\kappa$ is the signal field coupling and $\Gamma\equiv \Delta-\mi\gamma\equiv |\Gamma|e^{-\mi\theta}$ is the complex detuning, with real phase $\theta$. $\gamma$ arises from dephasing processes, including spontaneous emission. We have not included the Langevin noise operator which formally accompanies these loss terms, since its contribution vanishes when normally ordered expectation values of $A$ and $B$ are taken. We neglect the slow decay of the spin wave over the short timescale of the memory interaction.

We now introduce a new set of scaled coordinates: the \emph{memory time} $\epsilon(\tau)\equiv C\omega(\tau)/\omega(T)$, where $T$ is the duration of the interaction, and the \emph{effective distance} $\zeta(z)\equiv Cz/L$, with $L$ the length of the ensemble. Here $\omega(\tau)\equiv \int_0^\tau |\Omega(\tau')|^2\,\rmd \tau'$ is the \emph{integrated Rabi frequency} and $C\equiv|\kappa|\sqrt{L\omega(T)}/|\Gamma|$ is a coupling parameter. We define dimensionless annihilation operators $\alpha(\epsilon,\zeta)\equiv\sqrt{\omega(T)/C}e^{-\mi\chi(\tau,z)}A(\tau,z)/\Omega(\tau)$ for the optical field (assuming $\kappa$ is real for simplicity) and $\beta(\epsilon,\zeta)\equiv \sqrt{L/C}e^{-\mi\chi(\tau,z)}B(\tau,z)$ for the spin wave, where the exponent is $\chi(\tau,z)\equiv [\omega(\tau) +|\kappa|^2z]/\Gamma$. The first term in $\chi$ describes a Stark shift due to the control field; the second represents a modification of the signal group velocity. With these changes, the equations of motion reduce to the simple coupled system $\partial_\zeta\alpha=e^{\mi\theta}\beta$; $\partial_\epsilon \beta = -e^{\mi\theta}\alpha$. The solution of these equations then holds for all control pulse shapes and arbitrary inputs. The coupling parameter $C$ sets the size of the region in $(\epsilon,\zeta)$-space over which the memory interaction is driven. For the case considered here, an atomic ensemble, $C$ can be re-written in the form $C=(\pi \alpha_f \hbar/m_e)^{1/2}f\sqrt{N_aN_c}/(|\Gamma|\mathcal{A})$, where $f$ is the geometric mean of the oscillator strengths for the signal and control transitions, $\mathcal{A}$ is the cross-sectional area of the control field, and $N_a$ $(N_c)$ is the number of atoms (photons) interacting with (comprising) the control pulse. Here $\alpha_f$ is the fine structure constant, and $m_e$ is the electron mass.
 
The memory read-in and read-out must be unitary to function correctly. We now show that canonical evolution of the field operators $\alpha$ and $\beta$ is guaranteed by the classical structure of Eqs.~(\ref{MBs2.1},\ref{MBs2.2}) in the dispersive limit $\Delta \gg \gamma$. In this case, the phase $\theta$ vanishes and the following continuity relation is satisfied: $\partial_\zeta \alpha^\dagger \alpha +
\partial_\epsilon \beta^\dagger \beta = 0$. Integration of this expression over a square in $(\epsilon,\zeta)$-space yields the
flux-excitation conservation condition
\begin{equation}
\label{flux2} N_\alpha(C) + N_\beta(C)=N_\alpha(0) + N_\beta(0),
\end{equation}
where the number operators
$N_\alpha(\zeta)\equiv \int_0^C \alpha^\dagger(\epsilon,\zeta)\alpha(\epsilon,\zeta)\,\rmd \epsilon$, and $N_\beta(\epsilon)\equiv \int_0^C
\beta^\dagger(\epsilon,\zeta)\beta(\epsilon,\zeta)\,\rmd \zeta$, count the number of signal photons at an effective distance $\zeta$, and the number of
excitations of the spin wave at memory time $\epsilon$, respectively. Eq.~(\ref{flux2}) must hold for
arbitrary initial amplitudes $\alpha_0(\epsilon)\equiv\alpha(\epsilon,0)$ and $\beta_0(\zeta)\equiv\beta(0,\zeta)$, which fixes the transformation
$\left\{\alpha_0(\epsilon),\beta_0(\zeta)\right\}\rightarrow \left\{\alpha_C(\epsilon),\beta_C(\zeta)\right\}$ as unitary (where $\alpha_C(\epsilon)\equiv
\alpha(\epsilon,C)$ is the signal amplitude at the exit face of the ensemble, and $\beta_C(\zeta)\equiv\beta(C,\zeta)$ is the spin wave amplitude at the end of the read-in process). This allows the
dynamics to be decomposed into a set of independent transformations between light-field and spin
wave modes \cite{Wojtek}\cite{braunstein}. In what follows we therefore concentrate on the dispersive limit, and consider the case $\Gamma\rightarrow \Delta$; $\theta\rightarrow 0$. The solution of the dynamical equations is expressed by the
scattering relations \cite{Raymer1,raymer2}
\begin{eqnarray}
\label{exitfacea} \alpha_C(\epsilon)=\int_0^C [G_1(\epsilon-x,C)\alpha_0(x)
+ G_0(C-x,\epsilon)\beta_0(x)] \rmd x;\\
\label{exitfaceb} \beta_C(\zeta)=\int_0^C  [G_1(\zeta-x,C)\beta_0(x) -G_0(C-x,\zeta)\alpha_0(x)]\rmd x.
\end{eqnarray}
The integral kernels are given by $G_0(p,q)\equiv J_0 \left(2\sqrt{pq}\right)$, and $ G_1(p,q)\equiv \delta(p) -
\Theta(p)J_1\left(2\sqrt{pq}\right)\sqrt{q/p}$, with the $n^{\mathrm{th}}$ Bessel function of the first kind
denoted by $J_n$, and where the Heaviside step function $\Theta$ ensures that the convolutions in Eqs.
(\ref{exitfacea},\ref{exitfaceb}) respect causality.

\begin{figure}
\begin{center}
\includegraphics[height=3.5cm]{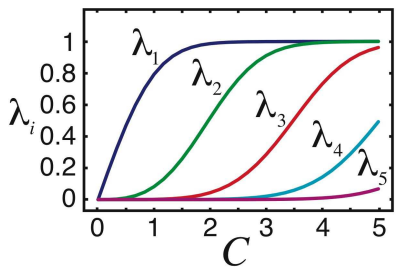}
\caption{The five largest singular values of the kernel $G_0$, plotted as a function of the coupling parameter
$C$.} \label{Figure2}
\end{center}
\end{figure}

\begin{figure}
\begin{center}
\includegraphics[height=3.5cm,width=\columnwidth]{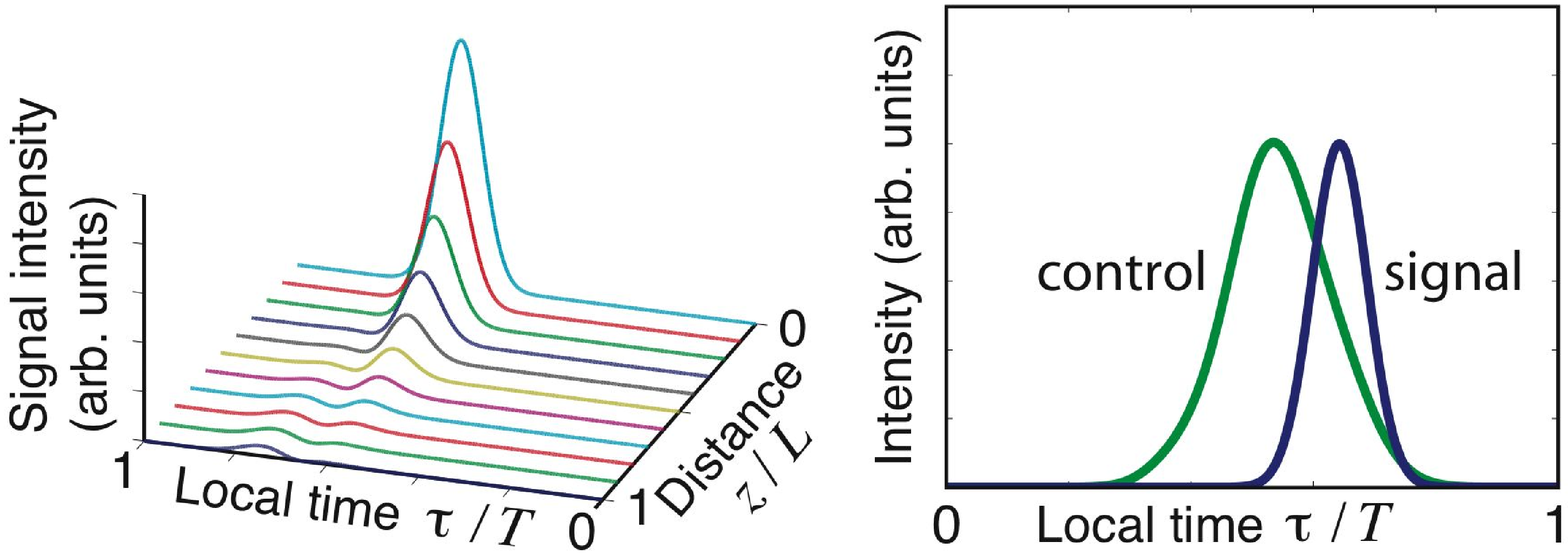}
\caption{Left: the intensity $\expect{A^\dagger(\tau,z)A(\tau,z)}$ of a Gaussian signal photon $\ket{1_\xi}$,
with wavepacket amplitude $\xi(\tau)\propto \exp{\{-2\ln 2[(\tau-\tau_0)/\sigma]^2\}}$, as it propagates through
an atomic ensemble with $C=2$. Here $\sigma=T/8$, $\tau_0=2T/3$. Right: the optimized control field intensity is
shown alongside the initial signal field intensity (scaled for clarity).} \label{Figure3}
\end{center}
\end{figure}

\begin{figure}
\begin{center}
\includegraphics[height=4cm]{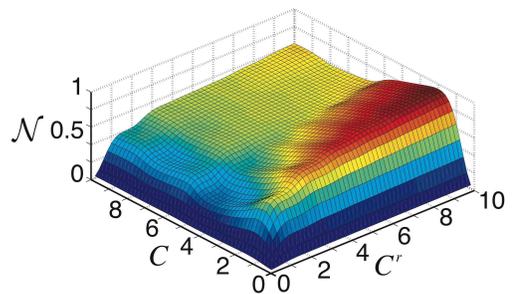}
\caption{The photon retrieval probability $\mathcal{N}$, for forward read-out, plotted as a function of the
read-in and read-out coupling parameters $C$ and $C^r$.} \label{Figure4}
\end{center}
\end{figure}

The integral kernels $G_{0,1}$ -- as they appear in Eqs.~(\ref{exitfacea},\ref{exitfaceb}) --
share symmetry under reflection about the line $C-x=y$, where $y$ stands for the
independent variable: either $\epsilon$ or $\zeta$. This symmetry, along with Eq.~(\ref{flux2}), allows us to decompose the
kernels using input and output modes related by time reversal (or equivalently space reversal) as follows:
\begin{eqnarray}
\label{simpleform0}
G_0(C-\epsilon,\zeta) =& \sum_{i=1}^\infty \phi_i(\zeta)\lambda_i \phi_i(C-\epsilon);\\
\label{simpleform1} G_1(\zeta-\epsilon,C) =& \sum_{i=1}^\infty \phi_i(\zeta)\mu_i \phi_i(C-\epsilon),
\end{eqnarray}
where $\{\phi_i\}$ is a complete orthonormal set of real mode\-functions and where the real, positive singular
values satisfy the constraint $\lambda_i^2 + \mu_i^2=1, \, \forall i$.
\paragraph{Read-in}
The ensemble begins the read-in process in the state $\ket{0}$, and we are free to replace $\beta_0(\zeta)$ by its expectation value; $\beta_0(\zeta)\rightarrow0$. With the above decomposition, Eq.~(\ref{exitfaceb}) then describes a mapping of the optical input mode $\phi_i(C-\epsilon)$ to the output spin wave mode $\phi_i(\zeta)$, with \emph{transfer amplitude} $-\lambda_i$, for each $i$. Transforming from memory time $\epsilon$ back to local time $\tau$, the normalized input modes are written $\Phi_i(\tau) \equiv \sqrt{C/\omega(T)}e^{\mi\chi(\tau,0)}\Omega(\tau)\phi_i[C-\epsilon(\tau)]$. The read-in efficiency can be quantified by evaluating the expectation value of the spin wave number operator $\expect{N_\beta(C)}$ at the end of the read-in process. Expanding an incident signal wavepacket $\xi(\tau)$ using the $\Phi_i$, the read-in efficiency is expressed as $\expect{N_\beta(C)}=\sum_i \lambda_i^2 |\xi_i|^2$, with the $i^{\mathrm{th}}$ \emph{overlap} given by $\xi_i\equiv \int_0^T \xi^*(\tau)\Phi_i(\tau)\,\rmd \tau$. When $\expect{N_\beta(C)}=1$, the read-in works perfectly. In Fig.~\ref{Figure2} the first five transfer amplitudes are plotted as a function of $C$.
These are found using the eigenvalue equation
\begin{equation}\label{eigenvals}
\int_0^C J_0(2\sqrt{xy})\phi_i(y)\,\rmd y = \lambda_i \phi_i(x),\end{equation} which we solve numerically using
a $500$ by $500$ square grid. It is desirable to limit the energy of the control pulse, so we should find the minimum coupling parameter $C$ which permits complete storage of the signal. For $C\approx 2$ the lowest mode achieves its optimal efficiency $\lambda_1\approx 1$, but higher modes remain poorly coupled. The efficiency of the memory is therefore maximized by setting $\xi_1= 1$; $\xi_{i\neq 1}= 0$, so that $\expect{N_\beta(C)}=\lambda_1^2\approx 1$ (for $C\geq 2$). To do this, it is necessary to shape the control field so that $\Phi_1(\tau)=\xi(\tau)$. If Gaussian optics ($\mathcal{A}\sim cL/\omega_s$) are used to illuminate a region a few cm long in a typical atomic vapour ($f\sim1$) of modest density ($\sim 10^{20}\,\mathrm{m}^{-3}$), with $100$ nJ control pulses, a $1$ ps photon wavepacket can be stored optimally, with $C=2$. In practice modematching could be achieved through measurement and feedback: the signal and control field sources are locked to a phase reference and operated in pulsed mode. The control pulse profile is characterized \cite{spider}, and augmented until the transmission of the signal is minimized. Figure ~\ref{Figure3} shows the result of a simple optimization to find the control pulse shape which modematches the lowest input mode to a Gaussian signal photon, for $C=2$. The photon is absorbed with $\expect{N_\beta(C)}\approx 0.96$; the small transmission probability is due to the limitations of our numerical modematching optimization.
\paragraph{Read-out}
Once a properly modematched photon has been read in to the quantum memory, the ensemble is left in the output
mode $\phi_1(\zeta)$, with probability amplitude $-\lambda_1$. We now consider the effect of sending a second
control pulse, propagating in the same direction as the initial control pulse, into the ensemble. The center
frequency, bandwidth, and intensity of this read-out pulse may differ from that of the first control pulse
(herein the read-in pulse). Let us use a superscript $r$ to indicate those quantities associated with the
read-out. We neglect decoherence and dephasing of the spin wave over the storage period and set $B^r(0,z)=B(T,z)$. This provides us with one boundary
condition; the second is that the signal field begins in its vacuum state at the start of the read-out process,
$\expect{N_\alpha^r(0)}=0$. The efficiency of the read-out depends upon the
degree to which the spin wave mode $\psi_1(z)\equiv \sqrt{C/L}e^{\mi\chi(T,z)}\phi_1(Cz/L)$ (written in terms of the ordinary spatial variable $z$)
overlaps with the input modes $\Psi_i^r(z)\equiv\sqrt{C^r/L}e^{\mi\chi^r(0,z)}\phi^r_i\left[C^r(1-z/L)\right]$ for the read-out process.
The functions $\{\phi_i^r\}$ solve the eigenvalue equation (\ref{eigenvals}), with $C$ replaced by $C^r$. A
measure of the efficiency of the memory is the expectation value of the output photon number operator
$\mathcal{N}\equiv\expect{N_\alpha^r(C^r)}=\lambda_1^2\sum_i \lambda_i^{r2}|f_i|^2$, with the read-out overlaps defined by $f_i \equiv \int_0^L\psi_1^*(z)\Psi_i^r(z)\,\rmd z.$ The parameter $\mathcal{N}$ is the probability of retrieving a
photon from the ensemble at read-out, given that a single modematched photon was sent in with the read-in pulse. If the detuning does not change too much, so that $|\Delta^r-\Delta|/\Delta\ll \Delta^r/(|\kappa|^2L)\sim\sqrt{N_c^r/N_a}$, then the phases $\chi$, $\chi^r$ of the spin wave modes approximately cancel, and then the stored spin wave is \emph{phasematched} to the readout modes.
In Fig.~\ref{Figure4} the variation of $\mathcal{N}$, under this approximation, is plotted as a function of the read-in and read-out coupling
parameters $C$ and $C^r$.
If $C=C^r$, then the lowest read-out mode is just the mirror image of the spin wave mode. For small $C$, the spin
wave mode is monotonic, and relatively flat; $f_1$ is therefore large. However the transfer amplitudes
$\lambda_1$ and $\lambda^r_1$ remain small, so the retrieval probability is low. Increasing $\lambda_1$ requires a larger $C$, but this produces a more asymmetric spin wave mode, and $f_1$ falls.
It is then necessary to increase $C^r$ above $C$, so that higher modes, with which the spin wave mode overlaps
significantly, are efficiently coupled to the optical field. The retrieval probability is maximized along the
line $C\approx 2$, which represents the optimal coupling for the read-in process. However, a read-out coupling
parameter in excess of $10$ is required to achieve $\mathcal{N}\geq 0.95$. Note also that modematching to the
lowest mode at read-in is the best strategy for maximizing the memory efficiency. Modematching to a higher mode, or some
combination of modes, simply increases the optimal read-in coupling above $C=2$ \footnote{This point will be clarified in a forthcoming article}.

The time reversal symmetry between the input and output modes makes the read-out for this scheme a non-trivial
problem. Simply repeating the read-in process (so that $C=C^r$) results in poor performance of the memory.
However, the dramatic increase in coupling strength required to extract the stored excitation fully, may make a
na\"{i}ve increase in control pulse energy at read-out prohibitively difficult to realize. An alternative method
to boost the coupling is to reduce the bandwidth of the read-out pulse, along with its detuning
$\Delta^r$. The photon recovered from the memory in this way would be frequency-shifted (according to the Raman
resonance condition), and temporally stretched (since its bandwidth would be diminished as well). Such a memory would act as a `photon transducer', storing broadband photons and converting them to narrowband photons with tunable frequency on demand.

Note that switching the propagation
direction of the read-out pulse sends $\phi_i(Cz/L)\rightarrow \phi_i\left[C(1-z/L)\right]$, so that the
spin wave mode overlaps exactly with the lowest read-out mode with $C^r=C$, and we should obtain
$\mathcal{N}=\lambda_1^4$. Unfortunately the read-out process is no longer phasematched in this situation, and the overlap integrals $f_i$ vanish. However, a solid-state implementation might allow this kind of reverse read-out with the use of \emph{quasi}-phasematching, in which the
sign of the read-out coupling parameter $C^r$ is periodically flipped along the length of the ensemble.

The phasematching
problem is obviated in the limit of vanishing Stokes shift, but then the ground state $\ket{0}$ must be prepared
artificially with high purity: if the state $\ket{3_i}$ is initially populated, or if selection rules allow
residual coupling of the control to the ground state, the memory fidelity at the level of single quanta is greatly reduced. Furthermore, if signal and control are spectrally indistinguishable, another degree of freedom should be used to differentiate between them. Typically, polarization affords discrimination to a part in $\sim 10^6$, but then the control should not contain more than a million photons. A large Stokes shift is therefore desirable, and correct phasematching at read-out is crucial for efficient retrieval. The above
considerations demonstrate the importance of propagation effects in the design of an optical quantum memory.
\paragraph{Summary}
We have shown that the off-resonant Raman configuration for a $\Lambda$-type ensemble quantum memory can be used to implement deterministic, controllable, and unitary transfer of the temporal structure of broadband single photons to a stationary spin wave, in the adiabatic regime. The dynamics are understood and optimized using a universal mode decomposition, valid for all control pulses and arbitrary input states. The modes are computationally simple to evaluate, and only a few of them are required to approximate the interaction faithfully. The optimal fidelity of the memory depends only upon a single dimensionless parameter ($C$), which defines an equivalence class for memories with different physical implementations.
\begin{acknowledgments}
The authors gratefully acknowledge the support of Hewlett-Packard and the EPSRC (UK) through the QIP IRC
(GR/S82176/01). MGR thanks the QIP IRC for support during his stay in Oxford, and also acknowledges the US NSF
(AMOP) for support. The research of DJ was supported in part by The Perimeter Institute for Theoretical Physics. IAW was supported in 
part by the European Commission under the Integrated 
Project Qubit Applications (QAP) funded by the IST 
directorate as Contract Number 015848. JN thanks A. V. Gorshkov for clarifying numerous points.
\end{acknowledgments}
\bibliography{references}
\end{document}